\documentstyle[12pt,aasms4]{article}
\makeatletter
\def\lnyoro{\mathrel{\mathpalette\gl@align<}}
\def\gnyoro{\mathrel{\mathpalette\gl@align>}}
\def\gl@align#1#2{\lower.6ex\vbox{\baselineskip\z@skip\lineskip\z@\ialign{$\m@th
#1\hfil##\hfil$\crcr#2\crcr\sim\crcr}}}
\makeatother

\begin{document}

\title{\bf Observational Test of Environmental Effects on The Local
Group Dwarf Spheroidal Galaxies}
\author{\bf Naoyuki Tamura$^{1}$ \& Hiroyuki Hirashita$^{1,2}$}
\affil
{$^1$Department of Astronomy, Faculty of Science, Kyoto
University, Sakyo-ku, Kyoto 606-8502, Japan}
\affil
{$^2$Research Fellow of Japan Society of the Promotion of Science}
\centerline{Aug. 10, 1999}
\centerline{email: tamura@kusastro.kyoto-u.ac.jp}
\authoremail{tamura@kusastro.kyoto-u.ac.jp}
\begin{abstract}

In this paper, we examine whether tidal forces exerted by the Galaxy or
M31 have an influence on the Local Group dwarf spheroidal galaxies
(dSphs) which are their companions. We focus on the surface brightness
profiles of the dSphs, especially their core radii because it is
suggested based on the numerical simulations that tidal disturbance can
make core radii extended. We examine the correlation for the dSphs
between the distances from their parent galaxy (the Galaxy or M31) and
the compactnesses of their surface brightness profiles by using a
parameter ``$C$'' defined newly in this paper.  Consequently, we find no
significant correlation. We make some remarks on the origin of this
result by considering three possible scenarios; tidal picture, dark
matter picture, and heterogeneity of the group of dSphs, each of which
has been often discussed to understand fundamental properties and
formation processes of dSphs.



\end{abstract}

\section{Introduction}

Recent observations have been revealing the physical properties of the
Local Group dwarf spheroidal galaxies (dSphs). The dSphs have
luminosities of order $10^5$--$10^7\, L_\odot$, and are characterized by
their low surface brightnesses (Gallagher \& Wyse 1994 for review).

The observations of such low-luminosity objects are important for
several reasons. One of them is that we can examine the environmental
effects in detail by the observational data because such objects with
small binding energies may be easily affected by their environments. For
the Local Group dSphs, the tidal forces exerted by the Galaxy or M31 are
likely to be the most important environmental effects. In fact, for
example, Kroupa (1997) and Klessen \& Kroupa (1998) theoretically
discussed the fate of dwarf satellite galaxies based on the tidal
effects, and Bellazzini et al. (1996) presented an observational support
by examining correlations between surface brightness and tidal force
(but see Hirashita, Kamaya \& Takeuchi 1999).

If the tidal forces really have major effects on the dSphs, a dwarf
galaxy closer to a giant galaxy (in the Local Group, the galaxy or M31)
should be more disturbed and have a more extended surface brightness
profile. In this paper, we independently examine this point from the
observational point of view by introducing a ``compactness parameter''
derived from the core radius of the surface brightness profile (\S 3.1). 
Since the new data of the companions of M31 have recently been available
(e.g., Armandroff et al. 1998; Caldwell 1999; Grebel \& Guhathakurta
1999; Hopp et al. 1999), we use these dSphs as well as those surrounding
the Galaxy.

This paper is organized as follows. First of all, in the next section,
we present the sample and data. Then, we introduce the ``compactness
parameter'' and present the result of our analysis in \S 3. In \S 4, the
dark matter problem for the dSphs is discussed based on the result in \S
3. Finally, we summarize the content of this paper in \S 5.

\section{Sample and Data}

The physical parameters ($V$ band absolute magnitude, core radius, and
galactocentric distance from the parent galaxy) of the Local Group dSphs
in Mateo (1998) are used, except for And V, VI, VII. We refer to
Caldwell (1999) for these three dSphs. The adopted quantities are
presented in Table 1. The galacticentric distances for the companions of
our galaxy are derived from their heliocentric distances (Mateo 1998 and
references therein). For the companions of M31, we calculate the
distances from M31 by using both their projected and heliocentric
distances, taking into account the distance from M31 to us (770 kpc;
Mateo 1998), and these are presented in the column of $R_{GC}$.

\section{Results}

\subsection{Definition of compactness parameter $C$}

The physical parameters (luminosity, radius, and velocity dispersion) of
the dwarf elliptical galaxies (dEs) and dSphs as well as normal
elliptical galaxies are known to correlate (e.g., Peterson \& Caldwell
1993). Since the dSph sample shows a more significant scatter in the
correlation than the other ellipticals (Caldwell et al. 1992), we
examine whether the scatter is caused by the environmental effect from
the parent galaxies. For this purpose, we define the ``compactness
parameter'' by utilizing the relation between core radius ($r_{c}$) and
$V$ band absolute magnitude ($M_{V}$). We present the data plotted on
the $\log M_{V} - \log r_{c}$ plane in Figure 1. The locus of dwarf
elliptical galaxies in Peterson \& Caldwell (1993) is also shown by the
dotted square marked with dEs. In the following, we present the
definition of the ``compactness parameter''.

First of all, we determine the standard core radius ($r_{c,0}$) for each
dSph from the following relation,
\begin{eqnarray}
\log r_{c,0} = a M_{V} + b.\label{eq:basic} 
\end{eqnarray}
Peterson \& Caldwell (1993) analyzed 17 dEs and found that there is the
scaling relation between their effective radii ($R_{e}$) and $V$ band
luminosities ($L_{V}$) as
\begin{eqnarray}
L_{V} \propto R_{e}^{5.0 \pm 0.5}.
\end{eqnarray}
Here, we use this scaling relation to obtain the constant ``$a$'' in the
equation (\ref{eq:basic}). Although we use core radii unlike Peterson \&
Caldwell (1993) (they used effective radii), this has little effect on
the following result since there is only a small difference between core
radii and effective radii (e.g., Caldwell 1999).

Assuming that 
\begin{eqnarray}
L_{V} \propto r_{c,0}^{5.0},
\end{eqnarray}
we obtain the following relation between $r_{c,0}$ and $M_{V}$: 
\begin{eqnarray}
\log r_{c,0} = -0.080 M_{V} + 1.42.
\end{eqnarray}
We adopt the zero point ``$b$'' so that the averaged values of $\log
r_{c}$ and $M_{V}$ for our sample galaxies
(\verb|<|$\log r_{c}$\verb|>| $ = 2.34$, \verb|<|$M_{V}$\verb|>| $ = -
11.6$) satisfy the above equation, though the way to determine the zero
point does not matter to the following analysis. Note that we use
$r_{c,0}$ to indicate a core radius obtained for each dSph by
substituting the observed $M_{V}$ of the galaxy into the above mean
relation. Finally, we define the ``compactness parameter'' ($C$) as
\begin{eqnarray}
C \equiv \log(r_{c}/r_{c,0}),
\end{eqnarray}
where the values of $r_{c}$ are listed in Table 1. Here, we comment on
an error of $C$ (referred to $\Delta C$ hereafter), which is determined
from an error of $r_{\rm c}$. Since errors of $r_{\rm c}$ are presented
in Mateo (1998), we find that almost all the absolute values of $\Delta
C$ are smaller than 0.1 as given in the column of $C$ in Table 1. It
should be noted that most of the errors of $\log R_{\rm GC}$ are smaller
than 0.1 as seen in Table 1. These errors are small enough to make our
following discussions valid. If $r_{c}$ is larger than $r_{c,0}$, in
other words, $C > 0$ for a dSph, we should consider that the galaxy is
more extended than it should be for its luminosity. In the context of
the environmental effect, we could find a negative correlation for the
sample dSphs between $C$ and $R_{\rm GC}$ since a dSph closer to a giant
galaxy should be more disturbed and more extended by the tidal effects.

\subsection{Result}

We present the data of our sample dSphs plotted on the $\log R_{\rm GC} -
C$ plane in Figure 2. There, the Galaxy's companions are indicated by
filled squares, and M31's by open squares. The correlation coefficient
between $\log R_{\rm GC}$ and $C$ with all the data is $- 0.28$.
Dividing our sample into the Galaxy's companions and the M31's ones, the
coefficients become $-0.36$ and $+0.12$, respectively. That is, we find
no significant correlation. Note that this conclusion is not altered
even if a different inclination of the $\log r_{\rm c,0} - M_{V}$
relation ($a$ in eq.1) is adopted within a reasonable range. Although a
mean value of $C$ for the M31 companions may be smaller than that for
the Galaxy's, this difference is not significant considering a large
scatter of $C$. For the Galaxy's companions, Sculpter could be seen as
an exception and if it is removed from them, there might be the
correlation (the correlation coefficient could be $- 0.66$). This may
suggest that the group of so-called dSph satellites is heterogeneous, as
discussed in \S 4.2.

\section{Discussion}

\subsection{The dark matter problem}

To date, stellar velocity dispersions of dSphs have been extensively
measured (e.g., Mateo et al. 1993 and references therein), which, in
general, indicate too large mass to be accounted for by the visible
stars in the dSphs. In other words, dSphs have generally high
mass-to-light ratios. This fact may imply the presence of dark matter
(DM) in these systems (e.g., Mateo et al. 1993). Existence of DM is
supported by the large spatial distribution of stars to their outer
regions (Faber \& Lin 1983) and the relation between the physical
quantities of the dSphs (Hirashita et al. 1999, but see Bellazzini et
al. 1996). Moreover, on the basis of this DM picture, the relation
between the ratio of the virial mass to the $V$-band luminosity and the
virial mass (the $M_{\rm vir}/L - M_{\rm vir}$ relation) for the Local
Group dSphs is naturally understood as the sequence of their star
formation histories in their forming phases by quasistatic collapse in
the DM halo (Hirashita et al. 1998).

However, the above arguments may be challenged if we consider the tidal
force exerted by the Galaxy.  If a dwarf galaxy orbiting a giant galaxy
(the Galaxy or M31 in the Local Group) is significantly perturbed by the
tides of the giant galaxy, the observed velocity dispersion of the dwarf
galaxy can be larger than the gravitationally equilibrium dispersion
(Kuhn \& Miller 1989; Kroupa 1997). This tidal picture of the dSphs also
suggests that the large velocity dispersions do not necessarily show the
existence of DM. Indeed, Klenya et al. (1998) demonstrated that Ursa
Minor has a statistically significant asymmetry in the stellar
distribution which can be attributed to tidal effects.

In summary, about the large stellar velocity dispersions and the large
$M_{\rm vir}/L$ of the dSphs, two major models are possible; the tidal
heating without DM and the presence and dominance of DM. Although we
cannot give a clear answer as to which of these models has more
validity, we discuss this problem taking into consideration the result
obtained in \S 3.2.

\subsection{Remarks based on our result}

\subsubsection{Tidal picture}

From the absence of the correlation between $C$ and $\log R_{\rm GC}$ as
shown in \S 3.2, it is not suggested that the tidal forces have major
effects on the dSphs, irrespective of whether the resonant orbital
coupling (Kuhn \& Miller 1989) could occur or not. However, we emphasize
that once the sample is split into the well-studied Galactic satellites
on the one hand, and the satellites of M31 on the other hand, then the
values and behaviour of $C$ with $R_{\rm GC}$ are consistent with the
tidal forces being important, at least for the companions of the Galaxy. 
It is noted that an orbit of a satellite may not be circular. If the
orbits of the dSphs are elliptical, their present $R_{\rm GC}$'s might
not reflect their averaged $R_{\rm GC}$ from past to present and $R_{\rm
GC}$ may not a good measure of tidal effects unless $R_{\rm GC}$ is very
small or very large. On the other hand, a satellite can pass near the
parent galaxy frequently enough to allow serious tidal perturbation
within a Hubble time even if the semi-major axis is 100 kpc. Thus, the
correlation should disappear and the tidal picture might not be rejected
completely from our result. However, if we consider a dSph in the
elliptical orbit around a giant galaxy and is observed at a location
relatively far from the giant galaxy, the dSph could not experience the
galactic tide unless the orbit is highly elliptical. In this case, since
the duration staying around the apogalacticon should be much longer (\S
VI of Searle \& Zinn 1973), the galaxy would not suffer the tidal effect
from the giant galaxy enough to be disturbed. Moreover, it should be
noted that $R_{\rm GC}$s in our sample widely spread from $\sim 20$ to
$\sim 300$ kpc. If $R_{\rm GC}$ is $\sim 300$ kpc, the orbital period is
expected to be comparable to a Hubble time by assuming the Keplerian
motion, which is adopted because orbits of the dSphs are still unknown
in detail. Consequently, though some difficulties may exist, a dwarf
galaxy closer to a giant galaxy (in the Local Group, the Galaxy or M31)
should be more disturbed, and thus, it is likely that the correlation
appears. That the sample of M31 companions shows a smaller correlation
than the Galactic sample may be due to the fact that the
three-dimensional distance estimates M31-satellites are very uncertain,
and therefore that projection can hide the tidal signature for the
sample of M31 satellites.

\subsubsection{DM picture}

In the DM model, the DM in the formation epoch may have determined the
star formation efficiency (Hirashita et al. 1998) and also the present
physical state of the galaxy. Assuming that the masses of the dSphs are
dominated by the DM, the tidal forces could have no effect on the dSphs
because their tidal radii are larger than their core radii (Pryor 1996). 
Since there is no reason why more extended dSphs are closer to the
Galaxy or M31 in the DM model, it seems rather natural that no
significant correlation between $R_{\rm GC}$ and $C$ is found. In other
words, the physical conditions of the dSphs should be determined by
their DM contents, not by their environments. Thus, the DM model does
not break down even in front of our result.

\subsubsection{Possible heterogeneity of the dSph sample}

It should be noted that the group of so-called dSph satellites may be
heterogeneous. Some may well be evolved ``normal'' low-mass galaxies in
the sense that they contain dark matter and have a cosmological origin,
and some may be secondary satellites that formed during mergers of
gas-rich protogalactic clumps contain little DM and have a globular
cluster-like origin. The latter of these systems will not contain dark
matter, and will be significantly affected by tides while orbiting
around the larger parent galaxy. Moreover, remnants of these can be
long-lived and may fake domination by DM (Kroupa 1997, Klessen \& Kroupa
1998). Therefore, it is worthwhile examining whether our result make
some difficulties in accepting such heterogeneity. In Figure 2, all the
Milky Way satellites except for Sculpter, have positive values of $C$,
and in addition, their $\log R_{GC}$ and $C$ seems to correlate. Note
that Sagittarius is believed to be experiencing serious tidal
modification. Thus it may be true that among the satellites of the
Galaxy in our sample, only Sculpter is exceptionally a DM dominated
dSph, and the tidal effects are the dominating factor for the
others. However, since there is no clear evidence of the heterogeneity,
all we can do here is to mention it as one possibility.

\section{Summary}

In order to investigate the tidal effect on the Local Group dSphs, we
examined the correlation between the distances from their host galaxy
(the Galaxy or M31) and the compactnesses of their surface brightness
profiles, ``$C$'' defined newly in this paper for the dSphs. As a
result, we find no significant correlation and thus no direct evidence
that tidal effects have a major effect on the dSphs. However, in most
cases, $C$ is sufficiently large to allow the possibility of tidal
effects, especially so since $C$ decreases for the furthest dSph
satellites of the Galaxy. Based on this result, we discussed the
validity of the existing pictures which have been suggested to explain
fundamental properties, especially the origin of their large
mass-to-luminosity ratios, of the dSphs.


\acknowledgements 
We are grateful to the anonymous referee for many useful comments to
improve the paper very much. We thank Dr. K. Ohta and Dr. S. Mineshige
for continuous encouragement. One of us (H.H) acknowledges the Reserch
Fellowship of the Japan Society for the Promotion of Science for Young
Scientists.

\newpage

\begin{table}
 \caption{Parameters of the Local Group dwarf spheroidal galaxies}

 \begin{tabular}{cccccc}
  \hline\hline
   Galaxy Name & $M_{V}$ & $r_{c}$ 
               & $R_{\rm GC}$\tablenotemark{a} 
               & $C$\tablenotemark{b}
               & Parent Galaxy \\ 
               &  (mag)  &  (pc) &  (kpc) &       &           \\ \hline
   Sagittarius
   & $-13.4$ &  550\tablenotemark{c}   &   16$\pm2$  &    0.28
   &           \\ 
   Draco       &  $-8.8$ &  180$\pm43$ &   76$\pm6$  &    0.16$\pm0.09$
   &           \\
   Carina      &  $-9.3$ &  210$\pm29$ &   89$\pm5$  &    0.19$\pm0.05$
   &           \\
   Ursa Minor  &  $-8.9$ &  200$\pm15$ &   66$\pm3$  &    0.20$\pm0.03$
   &           \\
   Sextans     &  $-9.5$ &  335$\pm24$ &   91$\pm4$  &    0.38$\pm0.03$
   & the Galaxy\\
   Sculptor    & $-11.1$ &  110$\pm30$ &   78$\pm4$  & $-$0.24$\pm0.10$
   &           \\
   Fornax      & $-13.2$ &  460$\pm27$ &  133$\pm8$  &    0.22$\pm0.02$
   &           \\
   Leo I       & $-11.9$ &  215$\pm20$ &  270$\pm30$ & $-$0.01$\pm0.04$
   &           \\ 
   Leo II      &  $-9.6$ &  160$\pm33$ &  219$\pm12$ &    0.05$\pm0.08$
   &           \\ \hline
   NGC 147
   & $-15.5$ &  170\tablenotemark{c}   &  109$\pm22$ & $-$0.40
   &           \\ 
   NGC 185     & $-15.5$ &  155$\pm47$ &  178$\pm21$ & $-$0.44$\pm0.12$
   &           \\
   NGC 205     & $-16.6$ &  260$\pm5$  &   45$\pm35$ & $-$0.30$\pm0.01$
   &           \\
   And I       & $-11.9$ &  375$\pm19$ &   57$\pm30$ &    0.23$\pm0.02$
   &           \\
   And II      & $-11.1$ &  205$\pm10$ &  281$\pm88$ &    0.03$\pm0.02$
   &  M31      \\
   And III     & $-10.3$ &  180$\pm24$ &   67$\pm16$ &    0.04$\pm0.05$
   &           \\
   And V       &  $-9.1$ &  110$\pm5$  &  115$\pm23$ & $-$0.08$\pm0.02$
   &           \\
   And VI      & $-11.3$ &  286$\pm7$  &  280$\pm85$ &    0.16$\pm0.01$
   &           \\ 
   And VII     & $-12.0$ &  240$\pm6$  &  215$\pm75$ &    0.03$\pm0.01$
   &           \\ \hline\hline
 \end{tabular}

 \tablenotetext{a}{This value means the Galactocentric distance from
 each galaxy for the Galaxy's companion, and the distance from M31 for
 M31's companion.}
 \tablenotetext{b}{$C$ represents ``compactness parameter'' (see text in
 detail).}
 \tablenotetext{c}{Error bars of $r_{c}$ are not attached to in the
 literatures.}

\end{table}

\newpage
\centerline{\bf Figure Caption}

\noindent
{Fig. 1---}
The relation between the core radius ($r_{c}$) and $V$ band absolute
magnitude ($M_{V}$) for the sample dSphs is shown.  Filled squares
indicate the dSphs which are companions of the Galaxy and open squares
indicate those of M31. The solid line indicates the relation that we use
to derive the standard core radius for each dSph (see text for detail).
The area marked with dEs represents a typical locus of dwarf elliptical
galaxies (e.g., Peterson \& Caldwell 1993).

\noindent 
{Fig. 2---}
Log $R_{\rm GC} - C$ relation (see text for their detailed definitions)
of our sample dSphs. Filled squares and open squares have the same
meanings as those in Fig. 1.


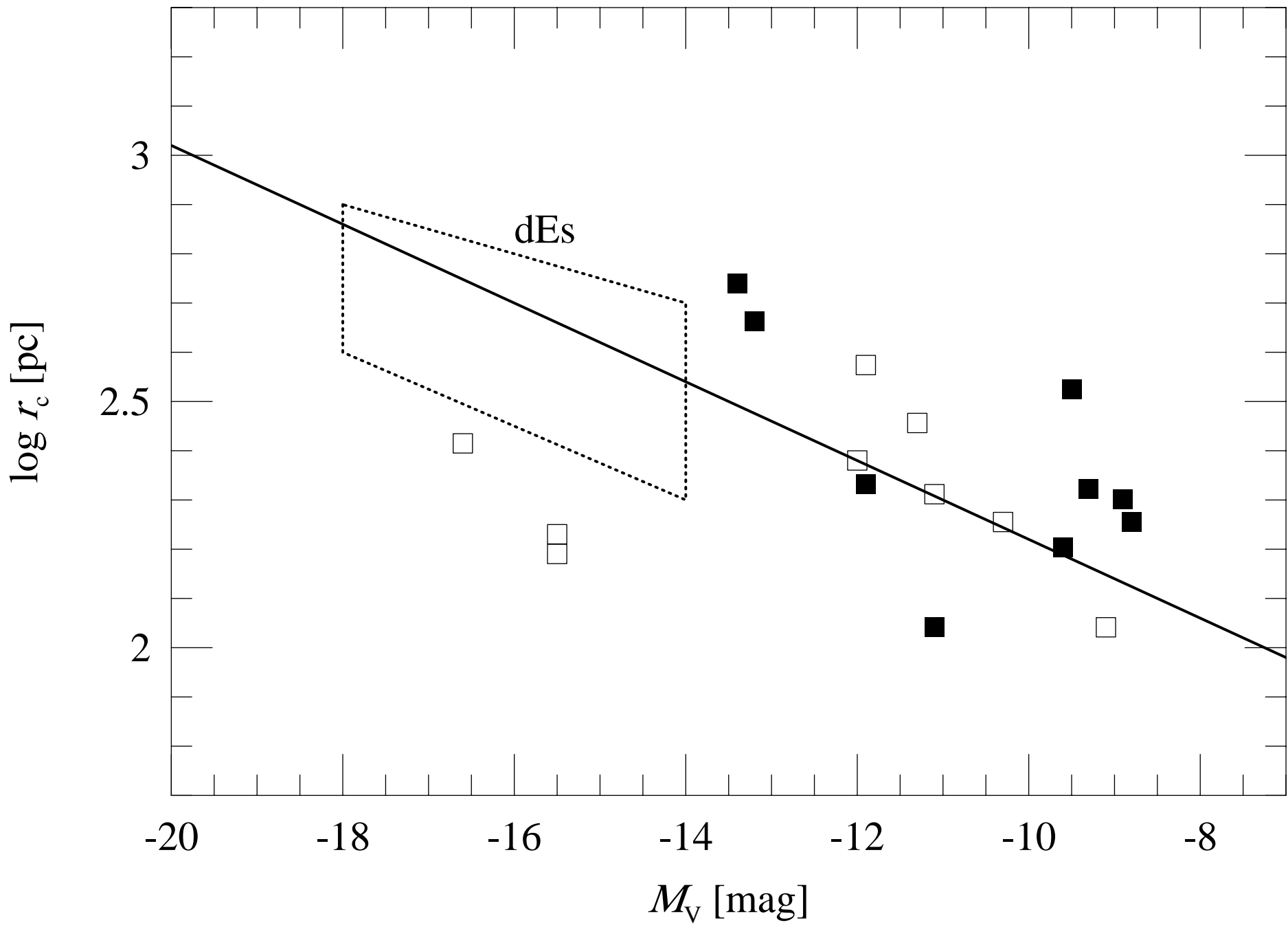
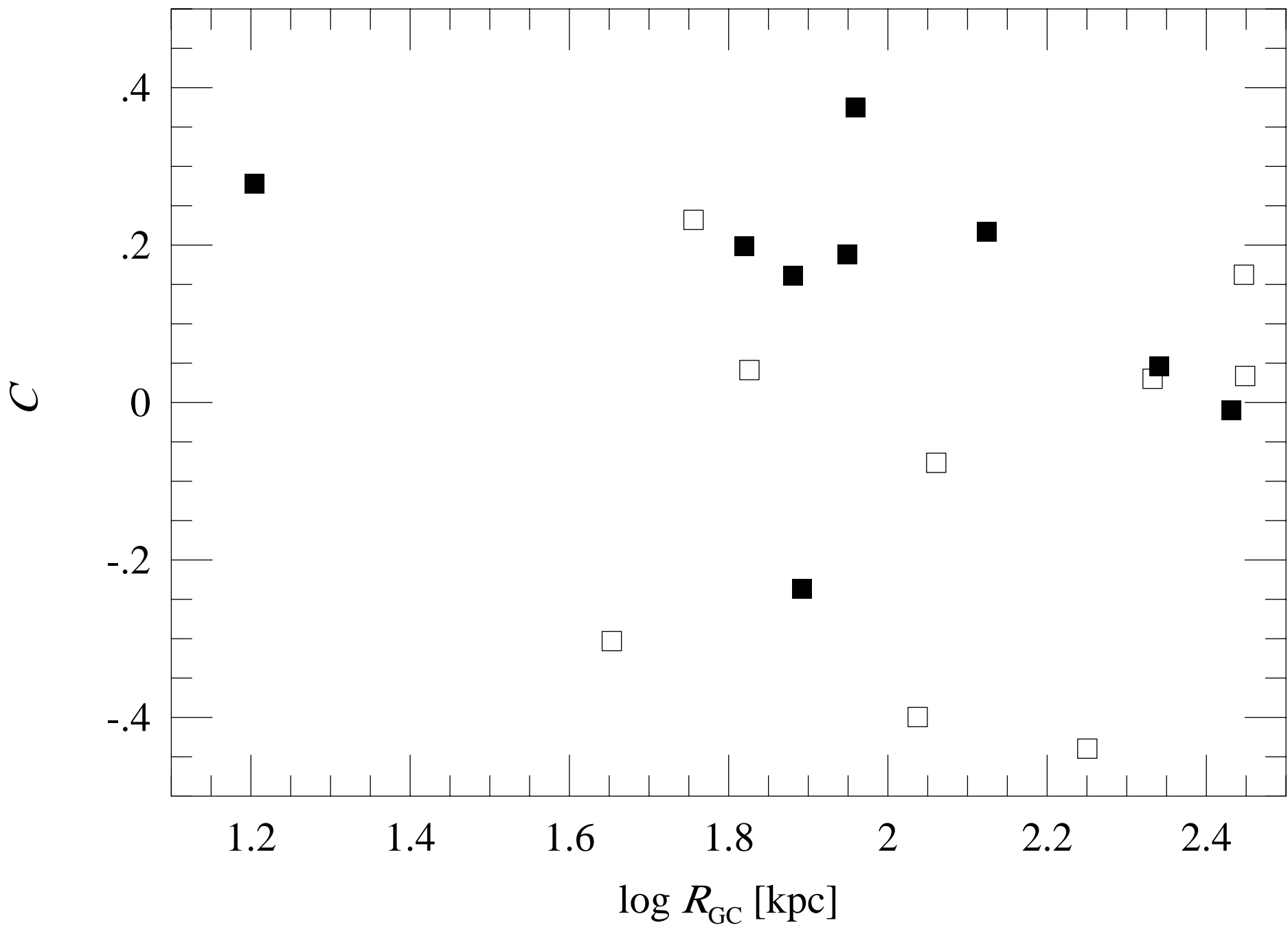
\end{document}